\documentclass[aps,pre,preprint,nofootinbib]{revtex4}
\usepackage{graphicx,xr,soul}

\begin{document}
\bibliographystyle{apsrev}

\title{Entropy and enthalpy of interaction between amino acid side chains in nanopores}

\author{S. Vaitheeswaran$^{1,3}$ and D. Thirumalai$^{1,2}$,\\
$^1$Biophysics Program, Institute for Physical Science and Technology,\\
$^2$Department of Chemistry and Biochemistry,\\
University of Maryland, College Park, MD 20742\\
$^3$\emph{Current address:} Department of Chemistry, University of Massachusetts,\\
Amherst, MA 01003\\
Email: {\tt vaithee05@gmail.com}}

\begin{abstract}
Understanding the stabilities of proteins in nanopores requires a quantitative description of  confinement induced interactions between amino acid side chains.
We use molecular dynamics simulations to study the nature of interactions between the side chain pairs  ALA-PHE, SER-ASN and LYS-GLU in bulk water and in water-filled nanopores.
The temperature dependence of the bulk solvent potentials of mean force and the interaction free energies in cylindrical and spherical nanopores is used to identify the corresponding entropic and enthalpic components.
The {\it entropically} stabilized hydrophobic interaction between ALA and PHE in bulk water is {\it enthalpically} dominated upon confinement depending on the relative orientations between the side chains.
In the case of SER-ASN, hydrogen bonded configurations that are similar in bulk water are thermodynamically distinct in a cylindrical pore, thus making rotamer  distributions  different from those in the bulk.
Remarkably, salt bridge formation between LYS-GLU is stabilized by entropy in contrast to the bulk. Implications of our findings for confinement-induced alterations in protein stability are briefly outlined.
\end{abstract}

\maketitle

Confinement effects are important in many biological processes such as chaperonin-assisted folding \cite{Thirumalai_Lorimer_ARBBS01}, polypeptide conformation in the exit tunnel of the ribosome
\cite{Woolhead_Johnson_Cell04}, translocation of peptides through biological channels \cite{Movileanu_Bayley_BiophysJ05},
and dynamics in the crowded cellular environment \cite{Minton_COSB00,Ellis_Minton_Nat03}. In addition, the effects of confinement on phase transitions in water-mediated interactions have significant applications in nanotribology, adhesion, and lubrication \cite{AlbaSimionesco06JPhysCondMatt}.
In light of their biological significance, several experimental \cite{Eggers_Valentine_JMB01,Eggers_Valentine_ProtSci01,
Ravindra_JACS04,Campanini_Bettati_ProtSci05,Bolis_JMB04}, theoretical and computational studies \cite{Betancourt_Thirumalai_JMB99,
Zhou_Dill_Biochem01,Klimov_Thirumalai_PNAS02,Ziv_Thirumalai_PNAS05,Cheung_Thirumalai_JMB06,Lucent_Pande_PNAS07,Zhou_JCP07,Mittal08PNAS,Wang09PNAS,Jewett04PNAS,Sirur13BJ}
have examined changes in the confinement-induced stability of proteins.
The interplay of several factors, such as alterations in hydrophobic and ionic interactions in
confined water, entropic restrictions of the conformations of polypeptide chains, and potential specific
interactions between amino acid residues and the confining boundary determine the stability of
proteins \cite{Betancourt_Thirumalai_JMB99,Cheung_Thirumalai_JMB06}.
A recent computational study \cite{Tian_Garcia_JACS11} showed that trp-cage confined in a fullerene ball is stabilized when the confining boundary is non-polar and destabilized when it is polar. Other computational studies have also reported stabilization of trp-cage \cite{Marino_Bolhuis_JPCb12} and a $\beta$-hairpin \cite{Bhattacharya_Mittal_BiophysJ12} by confinement between hydrophobic planar walls.
Despite the complexity, the observed enhancement in the stability of folded state, compared to that in bulk solvent \cite{Eggers_Valentine_JMB01,Ravindra_JACS04,Campanini_Bettati_ProtSci05,
Bolis_JMB04}, can often be rationalized in terms of the entropic stabilization theory
\cite{Betancourt_Thirumalai_JMB99,Zhou_Dill_Biochem01,Klimov_Thirumalai_PNAS02,Cheung_Thirumalai_PNAS05,Ziv_Thirumalai_PNAS05}.
However, confinement can also destabilize the folded state \cite{Cheung_Thirumalai_JMB06} due to the alterations in hydrophobic interactions, which can result in a net attraction between the protein and the confining boundary.
Destabilization of proteins upon confinement has been observed experimentally \cite{Eggers_Valentine_JMB01,
Eggers_Valentine_ProtSci01}, as well as in computer simulations \cite{Vaitheeswaran_PNAS08,Tian_Garcia_BiophysJ09}.

The interplay of a number of factors, including alterations in interactions between amino acid side chains upon confinement, determines changes in protein stability.
Previously, we showed \cite{Vaitheeswaran_PNAS08,Vaitheeswaran_JStatPhys11} that confinement in cylindrical pores greatly alters interactions between amino acid side chains (SCs): phenylalanine (PHE) - alanine (ALA), serine (SER) - asparagine (ASN) and lysine (LYS) - glutamate (GLU).
These SCs  are examples of large (PHE) and small (ALA) hydrophobic
species, polar (SER and ASN), and charged (LYS and GLU) moieties.
However, how confinement affects the entropy and enthalpy of interaction remains unknown. Computational models \cite{Baron_McCammon_JACS10,Setny_McCammon_JCTC10} have demonstrated that the thermodynamic signature of ligand binding to a receptor can vary from that for the association of molecular scale solutes in bulk water.  While molecular association in bulk water is driven by entropy, ligand binding to a pocket is enthalpy driven.
In order to provide further insights into the thermodynamics of protein folding in confined spaces, we report here,
the entropy and enthalpy of interaction between the three SC pairs in a cylindrical nanopore using molecular simulations. We use the temperature dependence of the potential of mean force to obtain the entropic and enthalpic contributions. Surprisingly, we find that interactions between PHE and ALA, which is entropically controlled in the bulk, is enthalpically stabilized upon confinement. In sharp contrast, salt-bridge formation is entropically stabilized in the nanopores whereas in the bulk the stabilization is dominated by enthalpy. These findings have profound implications for folding in confined spaces, which we briefly outline.
 \begin{figure}[htbp]
   \centerline{\includegraphics[width=0.5\textwidth]{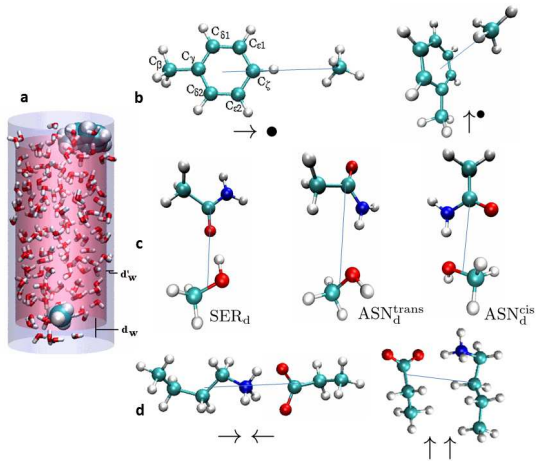}}
   \caption{(a) A snapshot of PHE and ALA in the $\uparrow^\bullet$ configuration in the cylindrical nanopore. Water is at bulk density and the side chains are pinned to the hydrophobic walls. Solvent-depleted zones are shown in light blue. (b) PHE and ALA in the $\rightarrow \bullet$ and the $\uparrow^\bullet$ orientations, (c) ASN and SER in the SER$_d$, ASN$_{\rm{d}}^{\rm{trans}}$ and ASN$_{\rm{d}}^{\rm{cis}}$ orientations (d) LYS and GLU in the $\rightarrow \leftarrow$ and $\uparrow \uparrow$ orientations. In (b-d), the reaction coordinate for each pair is the distance between their centers-of-mass, and is shown as a solid line.}
   \label{fig:structures}
 \end{figure}

 \begin{figure}[htbp]
   \centerline{\includegraphics[width=0.55\textwidth]{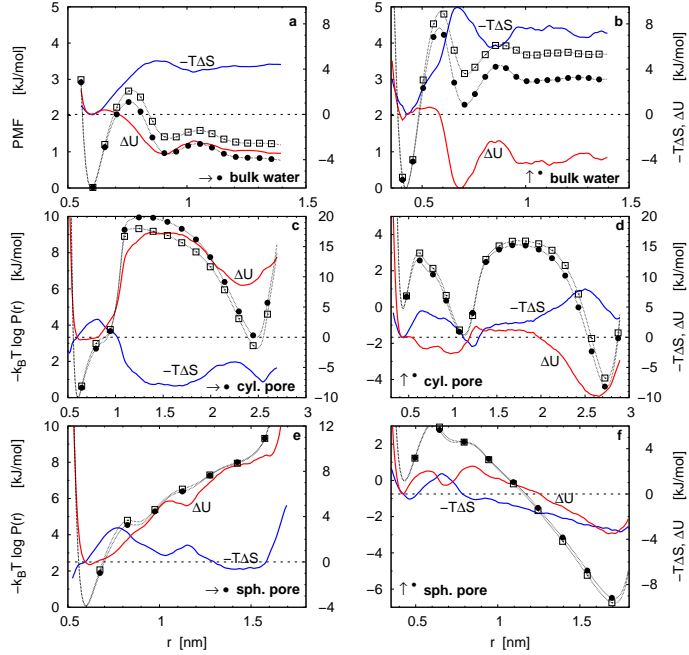}}
   \caption{ Temperature dependence of ALA-PHE PMFs in bulk water in the (a) $\rightarrow \bullet$ and (b) $\uparrow^\bullet$ orientation.  (c) and (d) show the interaction free energies in the cylindrical pore, while (e) and (f) are in the spherical droplet, respectively. In (a-f), solid circles correspond to 298 K and open squares to 328 K. The scale for the PMFs/free energies is on the left Y-axis; the right Y-axis shows the scale for the entropic (blue) and enthalpic (red) contributions.}
   \label{fig:Ala_Phe_multi}
 \end{figure}

\section{Methods}
{\it Systems:} For the ALA-PHE pair, the orientations sampled are:
(a) with the C$^{\rm{ALA}}$ atom in the same plane as the phenylalanine ring and closest to
the C$_{\zeta}^{\rm{PHE}}$ atom (which is the ring carbon that is the farthest from the
C$_{\beta}^{\rm{PHE}}$ atom), and (b) with the line joining the C$_{\zeta}^{\rm{PHE}}$ and
C$^{\rm{ALA}}$ atoms being perpendicular to the plane of the phenylalanine molecule (see Fig.~\ref{fig:structures}b).
These two orientations are denoted by the symbols $\rightarrow \bullet$ and $\uparrow^{\bullet}$ respectively
(The arrow points from C$_{\gamma}^{\rm{PHE}}$ to C$_{\zeta}^{\rm{PHE}}$).

The three hydrogen bonded orientations of the SER-ASN pair considered are (a) the serine -OH group as the H-bond donor and the asparagine carbonyl oxygen as the acceptor, (b) the serine oxygen being the
H-bond acceptor and the asparagine nitrogen and the trans hydrogen being the H-bond donor and, (c) same as (b) except the trans hydrogen is replaced by the cis hydrogen on the asparagine nitrogen.
In ASN, we label the hydrogen on the same side of the amide group as the carbonyl oxygen as cis, and the one on the opposite side as trans.
We refer to these as SER$_{\rm{d}}$, ASN$_{\rm{d}}^{\rm{trans}}$ and
ASN$_{\rm{d}}^{\rm{cis}}$, respectively (Fig.~\ref{fig:structures}c).

For the oppositely charged LYS$^+$-GLU$^-$ pair (Fig.~\ref{fig:structures}d), we consider
(a) the charged ends adjacent to each other and the linear SCs being colinear with each other
($\rightarrow \leftarrow$) and, (b) the charged ends adjacent to each other and the aliphatic
parts of the two SCs being parallel to each other ($\uparrow \uparrow$).
Thus, interactions between the $\rightarrow \leftarrow$ pair are mostly electrostatic while the
$\uparrow \uparrow$ pair will have strong electrostatic and hydrophobic interactions.

{\it Molecular Dynamics Simulations:} We study the interactions between the residue pairs ALA-PHE, SER-ASN and LYS-GLU using molecular dynamics simulations with the CHARMM22 force field \cite{MacKerell_Karplus_JPCb98} and TIP3P water
\cite{Jorgensen_Klein_JCP83}. In each amino acid, only the side chain (SC) is retained. The backbone atoms are deleted and the C$_{\alpha}$ atom is replaced by a hydrogen atom.
For instance, the side chain of ALA is represented by methane and PHE by toluene.
The interactions between the side chains (SCs) will vary with their relative orientations.
However, such multi-dimensional PMFs are prohibitively expensive to compute.
Instead, we sample $2-3$ representative orientations for each SC pair (Figs.~\ref{fig:structures}b-d), with the distance between their centers of mass as the reaction coordinate.
Relative orientations in each SC pair were maintained using a combination of angle and dihedral potentials on selected atoms.

In the unconfined systems, each pair of side chains (SCs) is solvated in approximately
807 water molecules with periodic boundaries, in a cubic cell about 3.0 nm in length.
Pressure is maintained at 1 bar using a Nos\'e-Hoover Langevin piston \citep{Martyna_Klein_JCP94,Feller_Brooks_JCP95}, and a temperature of 298 K or
328 K is set by Langevin dynamics.
Electrostatic interactions are evaluated using the particle mesh Ewald (PME) method.
We calculated the PMFs  using the adaptive biasing force (ABF) technique
\cite{Darve_Pohorille_JCP01,Henin_Chipot_JCP04} implemented in NAMD 2.6 \cite{Phillips_Schulten_JCC05}, in which the force acting along the reaction coordinate
(chosen to be the distance between the solute centers of mass) is evaluated and progressively refined during the
simulation. A biasing force is applied to counter this, so that the net force along the reaction coordinate
is zero. The mean force obtained from this process is integrated to give the corresponding potential.
PMFs and thus the free energies are only determined to within an additive constant; the value at contact
is arbitrarily set to zero.
The calculations were considered to be converged if the profiles did not change appreciably with increased
sampling. The convergence of an ABF calculation can also be verified by the uniformity of the sampling and
reversible diffusion of the system along the reaction coordinate.
Simulation times ranged from $70-400$ ns in bulk and $400-3600$ ns in the confined systems,
with integration time steps of 2 fs.
All bonds involving hydrogen atoms were frozen.
Data were accumulated over 10 independent trajectories for each case. The duration of the trajectories and the number of independent simulations are sufficient to obtain converged results.

{\it Creation of nanopores:} Cylindrical nanopores enclosing the solutes are carved out of the cubic cell after equilibration,
ensuring that the solvent inside is close to bulk density.
The solutes and solvent are confined by purely repulsive walls defined by the following potentials
in the cylindrical polar coordinates $\xi$ and $z$
\begin{eqnarray}
U_1(\xi) &=& 0 \hspace{5cm} \xi \le R \nonumber \\
       &=& \frac{k}{2}(R-\xi)^2 \hspace{3.4cm}  \xi >R \hspace{2cm}\rm{and}\\
U_2(z) &=& 0 \hspace{5cm} |z| \le 0.5L \nonumber \\
       &=& \frac{k}{2}(0.5L-z)^2 \hspace{2.9cm}  |z|>0.5L
\end{eqnarray}
where $R$ and $L$ are the pore radius and length respectively, and $k=83.6$ kJ/mol-\AA$^2$ in all cases.
The cylindrical pore is centered at the origin with its axis parallel to the z-axis, and has a diameter $D=2R=1.4$ nm and length $L=2.9$ nm.
In addition, the hydrophobic ALA-PHE pair is also simulated in a spherical water droplet of diameter $D=2R=2.0$ nm with a similar bounding potential.
Confinement potentials for the cylindrical pores and spherical droplets were implemented with cylindrical harmonic and spherical harmonic boundary conditions respectively, in NAMD 2.6 \cite{Phillips_Schulten_JCC05}.
Periodic boundary conditions are not applied and Lennard-Jones and electrostatic potentials are evaluated without a cutoff.

{\it Enthalpy and entropy of interaction:}  Interaction free energies, $-k_BT \log P(r)$ ($k_B$ is Boltzmann's constant , $r$ is the distance between the
centers of mass of the SC pair, and $P(r)$ is the probability of finding the two solutes at a separation $r$),
are calculated at a fixed volume and two temperatures, 298 K and 328 K, using ABF.
Because of the quasi one-dimensional nature of the confinement when $r \gg D$, we do not subtract
the free energy contribution $-2k_BT \log r$ which arises from the increase in phase space
proportional to $r^2$ in spherically symmetric systems.
Therefore, these profiles cannot be directly compared to potentials of mean force (PMFs) in bulk.
We evaluate the entropic and enthalpic components of the PMFs and interaction free energies assuming a
linear dependence on the temperature. The entropy change along the reaction coordinate
is given by $\Delta S(r) = - \partial w(r)/ \partial T,$ where $w(r)$ is the PMF/free energy.
The enthalpic contribution to   $w(r)$ is calculated as $\Delta U(r) = w(r) + T\Delta S(r)$.


{\bf Results}

{\bf Free energies of interaction between alanine and phenylalanine:}
We first discuss the thermodynamics of alanine-phenylalanine association.
Figs.~\ref{fig:Ala_Phe_multi}a and b show the PMFs for this pair in bulk water for
two orientations at two different temperatures.
The figures also show the enthalpic contribution to the PMF, $\Delta U(r)$, and the entropic
component, $-T \Delta S(r)$.
As expected for hydrophobic molecules \cite{Chandler_Nat05,Rick_Berne_JPCb97,Paschek_JCP04},
$-T \Delta S(r)$ drives the pair to contact, irrespective of their relative orientation.
$-T \Delta S(r)$ also gives rise to the desolvation barrier between the contact minimum and the solvent separated minimum (SSM) in the PMFs.
On the other hand, $\Delta U(r)$ favors extended separations, which is the source of the SSMs.

Confinement alters solvent-mediated ALA-PHE interactions.
Figs.~\ref{fig:Ala_Phe_multi}c and d show the interaction free energies in the cylindrical pore for this pair in both orientations.
When confined in the pore, the nonpolar solutes are hydrophobically adsorbed on to the walls.
Water hydrogen bonds are broken adjacent to the pore walls, and hence the solutes experience a different solvent environment than in the bulk.
The SSM, which arises due to a single intervening water molecule that is hydrogen bonded to other solvent molecules, is therefore completely absent in confinement \cite{Vaitheeswaran_JACS06,Vaitheeswaran_PNAS08}.
In the $\rightarrow \bullet$ orientation (Fig.~\ref{fig:Ala_Phe_multi}c), the contact state is
entropically {\it un}favorable compared to those at larger separations.
i.e. $-T \Delta S(r)$ favors extended separations between the solutes.
The large free energy barrier ($\sim 10$ kJ/mol, or 4 k$_B$T at 298 K) between the contact minimum and
the distant minimum at 2.5 nm arises entirely due to $\Delta U(r)$, which dominates overall for this system.
This pair is frustrated between the imposed orientational restraints, the cylindrical confinement and the thermodynamic drive to minimize solvent exposure.
In the $\uparrow^{\bullet}$ orientation (Fig.~\ref{fig:Ala_Phe_multi}d), the free energy profiles show
distant minima at $r \approx D$ and $L$.
The entropic contribution, $-T \Delta S(r)$, stabilizes the contact minimum and both the distant minima,
while $\Delta U(r)$ strongly favors the distant minimum at $\sim 2.7$ nm.

In a previous study \cite{Vaitheeswaran_JCP09}, we showed that for methane molecules in cylindrical water-filled pores,
interaction free energy profiles are similar to those for the $\uparrow^{\bullet}$ pair found in this study,
with distant minima at $r \approx D$ and $L$.
The contribution to their interaction free energy from the solute translational entropy, $-T \Delta S^{\rm{A}}(r)$,
was calculated by treating the methanes as point hard spheres pinned to the pore surface.
The calculations \cite{Vaitheeswaran_JCP09} showed that $-T \Delta S^{\rm{A}}(r)$ has a minimum at $r \approx$ min\{$D,L$\},
and accounts for the broad curvature of the free energy profile.
Here, qualitatively similar arguments can be made for the $\uparrow^{\bullet}$ pair.
In this configuration, the orientational restraints and the tendency to minimize the hydrophobic area exposed
to the solvent, can both be satisfied when the solutes are at the pore surface.
The planar PHE will be preferentially oriented parallel to the walls of the pore, either along its length,
or along the flat end caps.
With the solutes pinned to the surface, the minimum at $r \approx D$ in the free energy and also in
$-T \Delta S(r)$ (Fig.~\ref{fig:Ala_Phe_multi}d), will be mostly due to the translational entropy of the solutes.
The solute entropy will be low at contact, or at $r \approx L$, when they are located at opposite ends of the pore.
Therefore, the minima in $-T \Delta S(r)$ at the extremes of $r$ in Fig.~\ref{fig:Ala_Phe_multi}d can be attributed
to the gain in solvent entropy.

Interaction free energies between ALA and PHE, in a spherical pore of diameter $D=2.0$ nm, are plotted in
Figs.~\ref{fig:Ala_Phe_multi}e and f.
A comparison of these figures with Figs.~\ref{fig:Ala_Phe_multi}c and d, respectively, shows how the interaction between the
solutes is strongly dependent on the pore geometry.
The free energy for the $\rightarrow \bullet$ pair (Fig.~\ref{fig:Ala_Phe_multi}e), shows a single minimum at contact.
Furthermore, the interaction is almost entirely enthalpic; the entropy difference is small, except for the maximum separating the contact and solvent separated minima.
At contact, both nonpolar solutes can be in their preferred, solvent-depleted region at the surface.
But at intermediate and large separations, one or both solutes will be exposed to water, thus reducing solvent entropy.
The $\uparrow^{\bullet}$ pair (Fig.~\ref{fig:Ala_Phe_multi}f), is driven either to contact or to $r \approx D$, with nearly
equal contributions from $\Delta U(r)$ and $-T \Delta S(r)$.

The maximum in $-T \Delta S(r)$ that separates the contact and solvent separated minima at $r \approx 0.85$ nm for the $\rightarrow \bullet$ pair in bulk water (Fig.~\ref{fig:Ala_Phe_multi}a) can also be discerned in the cylindrical pore (Fig.~\ref{fig:Ala_Phe_multi}c) and the spherical droplet (Fig.~\ref{fig:Ala_Phe_multi}e). Similarly, the maximum in $-T \Delta S(r)$ at $r \approx 0.65$ nm for the $\uparrow^{\bullet}$ pair in the bulk (Fig.~\ref{fig:Ala_Phe_multi}b) also occurs in spherical and cylindrical confinement (Figs.~\ref{fig:Ala_Phe_multi}d and ~\ref{fig:Ala_Phe_multi}f respectively).
However, the thermodynamic signatures of ALA-PHE interactions are significantly altered by confinement. In the cylindrical pore and the spherical droplet, these hydrophobes are adsorbed on the concave confining wall.  This system can be viewed as a special case of Dzubiella's thermodynamic model \cite{Dzubiella_JStatPhys11}, which predicts that the hydrophobic interaction of a convex solute with a fully complementary concave surface is dominated by enthalpy.

{\bf Serine-Asparagine interactions:}
 \begin{figure}[htbp]
   \centerline{\includegraphics[width=0.55\textwidth]{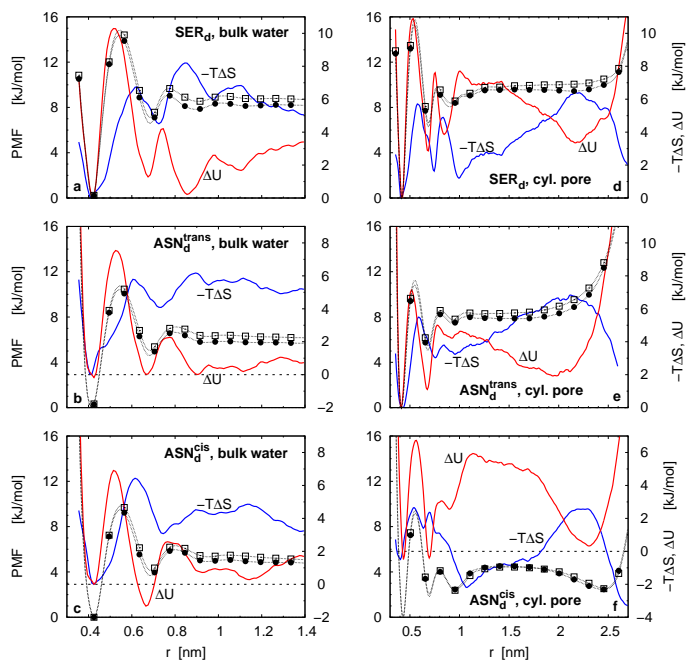}}
   \caption{Temperature dependence of PMFs of the SER-ASN pair in bulk water in the (a) SER$_{\rm{d}}$, (b) ASN$_{\rm{d}}^{\rm{trans}}$ and (c) ASN$_{\rm{d}}^{\rm{cis}}$ configurations. (d), (e) and (f) show the interaction free energies in the nanopore corresponding to (a), (b) and (c), respectively. In (a-f), solid circles correspond to 298 K and open squares to 328 K. The scale for the PMFs/free energies is on the left Y-axis; the right Y-axis shows the scale for the entropic (blue) and enthalpic (red) contributions.}
   \label{fig:Ser_Asn_multi}
 \end{figure}
The three different hydrogen bonding orientations of serine and asparagine in bulk water are
considered in Figs.~\ref{fig:Ser_Asn_multi}a-c.
The PMFs in bulk water in all three cases show a contact minimum and a SSM at $r \approx 0.7$ nm.
We see that the stability of the contact pairs, relative to extended separations at $r \approx 1.4$ nm, is mostly due to the favorable entropy, similar to the ALA-PHE pair.
In all three orientations, the desolvation barriers at
$r \approx 0.5$ nm are mostly enthalpic in origin, unlike the ALA-PHE pair.
Figs.~\ref{fig:Ser_Asn_multi}d-f show the corresponding interactions for the three pairs in cylindrical confinement.
For all three pairs, the second SSM at $r \approx 0.9$ nm is stabilized by confinement.
As noted before \cite{Vaitheeswaran_PNAS08}, confinement distinguishes the ASN$_{\rm{d}}^{\rm{trans}}$
and ASN$_{\rm{d}}^{\rm{cis}}$ configurations relative to each other.
In bulk solvent, these two configurations have similar interaction thermodynamics (Figs.~\ref{fig:Ser_Asn_multi}b and c).
However, in the pore, the ASN$_{\rm{d}}^{\rm{trans}}$ hydrogen bonded pair has significantly greater
stability than the ASN$_{\rm{d}}^{\rm{cis}}$ pair (Figs.~\ref{fig:Ser_Asn_multi}e and f).
The ASN$_{\rm{d}}^{\rm{cis}}$ orientation also has a distant minimum at $r \approx L$.
The solute pair is stabilized at this separation, at which the nonpolar methyl groups on SER and ASN
can be in their preferred environment at the pore surface away from water, while the polar ends
of both molecules remain hydrated.
In the pore, $-T \Delta S(r)$ favors the contact minimum for the SER$_{\rm{d}}$ and ASN$_{\rm{d}}^{\rm{trans}}$
pairs, but \emph{dis}favors the contact minimum for the ASN$_{\rm{d}}^{\rm{cis}}$ pair.
$\Delta U(r)$ stabilizes the contact minimum and both SSMs in all three orientations,
and is the major component of the free energy barriers at $r \approx 0.5$ and $\approx 0.8$ nm.

{\bf Lysine-Glutamate interactions:}
 \begin{figure}[htbp]
   \centerline{\includegraphics[width=0.55\textwidth]{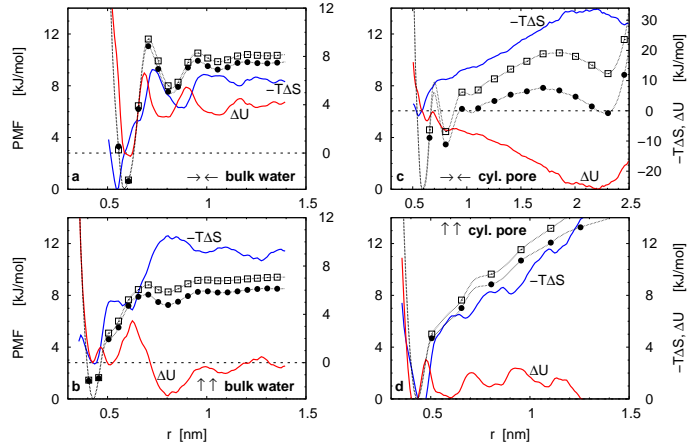}}
   \caption{Temperature dependence of PMFs of the LYS-GLU pair in bulk water in the (a) $\rightarrow \leftarrow$ and (b) $\uparrow \uparrow$ orientations.  (c) and (d) show the interaction free energies in the nanopore corresponding to (a) and (b), respectively. Solid circles correspond to 298 K and open squares to 328 K. The scale for the PMFs/free energies is on the left Y-axis; the right Y-axis shows the scale for the entropic (blue) and enthalpic (red) contributions.}
   \label{fig:Lys_Glu_multi}
 \end{figure}
The side chains of lysine and glutamate carry charges of +e and -e, respectively, at physiological pH.
They also have nonpolar parts that can interact hydrophobically.
Therefore the LYS$^+$-GLU$^-$ pair interaction can be mostly electrostatic or both electrostatic
and hydrophobic (amphiphilic), depending on their relative orientations.
The PMFs between this pair of side chains in bulk water is shown in Figs.~\ref{fig:Lys_Glu_multi}a and b.
For the $\rightarrow \leftarrow$ pair (Fig.~\ref{fig:Lys_Glu_multi}a), both $-T \Delta S(r)$ and
$\Delta U(r)$ drive the pair to contact.
On the other hand, the amphiphilic $\uparrow \uparrow$ pair (Fig.~\ref{fig:Lys_Glu_multi}b) is stabilized at
contact entirely due to the entropy.
$\Delta U(r)$ gives rise to the desolvation barrier between the contact minimum and the SSM, but contributes
little to the stability of the contact pair relative to large separations at $r > 1$ nm.
Figs.~\ref{fig:Lys_Glu_multi}c and d show the corresponding interaction free energies in the cylindrical pore.
The $\rightarrow \leftarrow$ orientation (Fig.~\ref{fig:Lys_Glu_multi}c) is strongly stabilized by the entropy.
Unlike the corresponding bulk PMF (Fig.~\ref{fig:Lys_Glu_multi}a), $\Delta U(r)$ strongly destabilizes the pair,
with a minimum at $\sim 2.2$ nm.
In the $\uparrow \uparrow$ orientation (Fig.~\ref{fig:Lys_Glu_multi}d), LYS and GLU are driven to contact entirely by entropy.
Similar to the bulk solvent case, $\Delta U(r)$ does not contribute to the stability of the contact pair.
Figs.~\ref{fig:Lys_Glu_multi}c-d show that confinement can stabilize salt bridges in proteins due to favorable entropic effects. This result is consistent with a recent computational study which found that, in a Lys-Glu dipeptide in a water cluster, the salt bridge is weakened as the cluster size increases \citep{Pluharova_JCP12}.

\bigskip

For all three solute pairs, ALA-PHE (Figs.~\ref{fig:Ala_Phe_multi}c, d), SER-ASN (Figs.~\ref{fig:Ser_Asn_multi}d-f)
and LYS-GLU (Figs.~\ref{fig:Lys_Glu_multi}c, d) in the cylindrical pore, $-T \Delta S(r)$ shows a distant minimum
at $r^m \approx L$. 
The corresponding $\Delta U(r)$ also has a minimum at a shorter separation $r^m_-$, where $r^m - r^m_-$
is characteristic of the molecular pair.
This minimum in $-T \Delta S(r)$ arises because solvent entropy is maximized when the solutes are at
opposite ends of the pore, with their nonpolar ends sequestered from the water.
Solute entropy will be low in this case.
As the inter-solute distance decreases from $r^m$ to $r^m_-$, $\Delta U(r)$ decreases due to
the gain in solute-solvent van der Waals interactions as the nonpolar ends are immersed in water.

\textbf{Conclusions:}
Water-mediated interactions between solute molecules in confinement differ drastically from those in bulk solvent.
From the temperature dependence of the bulk solvent PMFs and the interaction free energies in nanopores,
we have identified the corresponding entropic and enthalpic components.
Results for the ALA-PHE pair show that, while interactions between small hydrophobic molecules are indeed entropically
stabilized in bulk solvent \cite{Chandler_Nat05}, this is not necessarily the case in nanoporous confinement.
In confinement, the interaction thermodynamics depends on the balance between many factors, including the pore geometry,
the relative orientations of the interacting molecules and their preference for regions where solvent structure is disrupted.
When the relative orientations of nonpolar solutes are compatible with the pore geometry and their tendency for surface
solvation, the entropy of interaction will have a large contribution due to the solute translational entropy.
The polar pair of side chains SER-ASN, and the charged LYS-GLU pair have nonpolar parts that will be
sequestered at the pore surface where water hydrogen bonds are broken.
Therefore, the thermodynamics of association of SER with ASN, and LYS with GLU, are also altered by confinement.
Hydrogen bonded SER-ASN configurations that have similar thermodynamics in bulk water are thermodynamically distinct in a cylindrical pore.
Hence, rotamer population distributions are likely to be altered by confinement, compared to the bulk solvent case.
The LYS-GLU system also indicates that confinement is likely to entropically stabilize salt bridges in proteins.
Consequently, it is likely that the conformations sampled by a polypeptide chain in cavities, found for example in chaperonins, are likely to be very different from those in the bulk. This could lead to altered mechanisms for protein folding in confined spaces.

\section{Acknowledgement}
This work was supported by a grant from the National Science Foundation (CHE13-61946).


\end{document}